\documentclass[conference, a4paper]{IEEEtran} 

\usepackage{epsfig,graphics}
\usepackage{amssymb,amsfonts,latexsym}
\usepackage[cmex10]{amsmath}
\usepackage[usenames]{color}
\usepackage{array}
\usepackage{cite}
\usepackage{import}  
\usepackage[squaren]{SIunits}

\def\mathlette#1#2{{\mathchoice{\mbox{#1$\displaystyle #2$}}%
                               {\mbox{#1$\textstyle #2$}}%
                               {\mbox{#1$\scriptstyle #2$}}%
                               {\mbox{#1$\scriptscriptstyle #2$}}}}
\renewcommand{\Vec}[1]{\mathlette{\boldmath}{#1}}

\newcommand{\be}{\begin{equation}}
\newcommand{\ee}{\end{equation}}
\newcommand{\ba}{\begin{array}}
\newcommand{\ea}{\end{array}}
\newcommand{\bdm}{\begin{displaymath}}
\newcommand{\edm}{\end{displaymath}}
\newcommand{\bea}{\begin{eqnarray}}
\newcommand{\eea}{\end{eqnarray}}
\newcommand{\bean}{\begin{eqnarray*}}
\newcommand{\eean}{\end{eqnarray*}}
\newcommand{\me}{\text{e}}

\newcommand{\mj}{\text{j}}

\def\argmin{\mathop{\text{argmin}}}

\def\diag{\text{diag}}

\def\nTD{\ensuremath{T_\text{D}}} %delay spread
\def\nTC{\ensuremath{T_\text{C}}} %chip length
\def\nTS{\ensuremath{T_\text{S}}} %symbol length

\def\oH{\ensuremath{^\text{H}}} % Hermition operator
\def\oT{\ensuremath{^\text{T}}} % transpose operator
\def\nfC{\ensuremath{f_\text{C}}}        %carrier frequency
\def\nBD{\ensuremath{B_\text{D}}}        %Doppler bandwidth (one-sided)
\def\nEb{\ensuremath{E_\text{b}}}        %energy per information bit
\def\IDSFT{\ensuremath{\text{IDSFT}}}        %noise subspace

\IEEEoverridecommandlockouts

\title{Low-Complexity Equalization for Orthogonal Time and Frequency Signaling (OTFS)
} 

\author{
\IEEEauthorblockN{Thomas Zemen, Markus Hofer and David Loeschenbrand}
\IEEEauthorblockA{\textit{Security \& Communication Technologies, Center for Digital Safety \& Security}}
\textit{AIT Austrian Institute of Technology}\\
Vienna, Austria\\
\{thomas.zemen, markus.hofer, david.loeschenbrand.fl\}@ait.ac.at}

\begin{document}

\maketitle

\begin{abstract}
Recently, a new precoding technique called orthogonal time-frequency signaling (OTFS) has been proposed for time- and frequency-selective communication channels. OTFS precodes a data frame with a complete set of spreading sequences and transmits the results via orthogonal frequency division multiplexing (OFDM). OTFS uses two dimensional (2D) linear spreading sequences in time and frequency which are the basis functions of a symplectic Fourier transform. OTFS allows the utilization of time- and frequency-diversity but requires maximum likelihood decoding to achieve full diversity. In this paper we show performance results of a low-complexity equalizer using soft-symbol feedback for interference cancellation after an initial minimum-mean square error equalization step. Performance results for an implementation in the delay-Doppler domain and in the time-frequency domain are compared. With our equalizer, OTFS achieves a gain of 5dB compared to OFDM for a bit error rate of $10^{-4}$ and a velocity of $200\,\text{km/h}$.
\end{abstract}

\section{Introduction}

Reliable wireless communication links in non-stationary time- and frequency-selective channels are a key requirement for future 5G application scenarios such as connected autonomous vehicles, industry 4.0 production environments, and mm-Wave communication links (above 26 GHz).

Highly reliable wireless communication links require the utilization of all available diversity sources such as time, frequency, and spatial diversity. Orthogonal time frequency signaling (OTFS) is proposed in \cite{Rakib16, Monk16, Hadani17} and promises the full utilization of time- and frequency diversity.

Conventional orthogonal frequency division multiplexing (OFDM) \cite{Weinstein71} partitions the communication channel in orthogonal bins in the time and frequency domain, defining a two dimensional (2D) grid. OFDM transmits each data symbol on a single grid location in the time-frequency (TF) domain. OTFS uses a 2D discrete symplectic Fourier transform (DSFT) to precode the data symbols at the transmitter side. After this precoding operation the data is transmitted with conventional OFDM modulation. Hence, the information of each data symbol is linearly spread on all available grid points of a data frame in the TF domain. By appropriate equalization and decoding, the full time- and frequency-diversity of the wireless communication channel can be utilized on the receiver side. 

The precoding of data symbols in combination with OFDM was proposed by multiple authors in the context of multi-carrier (MC) code-division multiple access (CDMA) \cite{Linnartz01}. In the context of MC-CDMA the terms precoding and spreading are used interchangeably. In \cite{Svensson04} the authors discuss spreading in the frequency- (MC-CDMA), time- (MC direct sequences CDMA, MC-DSCDMA) or in both the time and frequency domain (time-frequency-localized, TFL-CDMA). In the context of MC-CDMA spreading is used to implement a multiple-access scheme and to obtain a  diversity gain. Iterative detection methods for MC-CDMA in time- and frequency selective channels are presented in \cite{Zemen06c, Dumard11}. In \cite{Debbah02} different complete bases for precoding in the frequency domain are analyzed.

The basic concept of OTFS is discussed in the white paper \cite{Monk16} and the patents published by Cohere \cite{Rakib16}. First performance results for OTFS and a comparison with OFDM is shown in \cite{Hadani17}. So far no information is available in the open literature on detailed equalization and detection algorithms.

OTFS encodes data symbols using orthogonal 2D basis functions. Their orthogonality is destroyed at the receiver (RX) side, due to the effect of the time- and frequency selective channel \cite{Verdu98b} as well as due to a possible windowing at the transmitter (TX) and RX side. This lost orthogonality leads to inter-symbol interference (ISI) of the data symbols contained in a frame. Hence, an appropriate equalization technique to remove the resulting ISI is required.

\subsection*{Contributions of the Paper:} 
\begin{itemize}
\item We include a matrix-vector formulation of OTFS to utilize algorithms from multi-user detection.
\item We present an ISI cancellation scheme by using soft-symbol feedback \cite{Zemen06c} to obtain a low-complexity maximum likelihood detection \cite{Dumard11} algorithm for OTFS. We discuss two implementations, one in the TF-domain and another one in the DSFT domain.
\item We compare the performance of both schemes by means of numerical link level simulations. Furthermore a comparison of OTFS with OFDM is presented. 
\end{itemize}

\subsection*{Notation:}
We denote a scalar by $a$, a column vector by $\Vec{a}$ and its $i$-th element with $a[i]$. Similarly, we denote a matrix by $\Vec{A}$ and its $(i,\ell)$-th element by $[\Vec{A}]_{i,\ell}$. The transpose of $\Vec{A}$ is given by $\Vec{A}\oT$ and its conjugate transpose by $\Vec{A}\oH$. A diagonal matrix with elements $a[i]$ is written as $\diag(\Vec{a})$ and the $Q\times Q$ identity matrix as $\Vec{I}_Q$. The absolute value of $a$ is denoted by $\left|a\right|$ and its complex conjugate by $a^*$. For the discrete set $\mathcal{I}$, $|\mathcal{I}|$ denotes the number of elements of $\mathcal{I}$. The Frobenius (2-norm) of a matrix or vector is denoted by $\|\Vec{A}\|$. We denote the set of all integers by $\mathbb{Z}$, the set of real numbers by $\mathbb{R}$ and the set of complex numbers by $\mathbb{C}$.

\vspace{1em}
\section{Signal Model}
\label{se:SignalModel}

OTFS \cite{Rakib16,Monk16, Hadani17} can use OFDM  as basic transport layer, i.e. the basic OFDM signal model  
\be
y[m,q]=w_\text{RX}[m,q]g'[m,q]w_\text{TX}[m,q]x[m,q]+n[m,q]
\label{eq:OFDMsignalmodel}
\ee 
applies. Here 
\be
m\in\{0,\ldots, M-1\}=\mathcal{I}_M
\ee
denotes discrete time, 
\be
q\in\{0,\ldots, N-1\}=\mathcal{I}_N
\ee
discrete frequency, $M$ the length of a data frame and $N$ the number of subcarriers. The sampled time-variant frequency response is denoted by $g'[m,q]$, the transmitted signal by $x[m,q]$, the received signal by $y[m,q]$ and additive white complex Gaussian noise with zero mean and variance $\sigma^2$ by $n[m,q]$. Equation \eqref{eq:OFDMsignalmodel} holds under two assumptions: First, the delay spread of the channel impulse response $\nTD$ must be shorter or equal than the length of the cyclic prefix
\be
\nTD\le G \nTC\,,
\ee
where $G$ denotes the length of the cyclic prefix in samples and $\nTC$ denotes the chip duration 
\be
\nTC=1/B
\ee
and $B$ the system bandwidth. Secondly, the Doppler spread $\nBD$ must be smaller than a fraction $\epsilon$ of the subcarrier bandwidth
\be
\nBD<\epsilon /(\nTC N)\,.
\ee 
A practical value for $\epsilon$ is $1\%$, i.e. $\epsilon=0.01$.

Optionally, 2D window functions $w_\text{TX}[m,q]$ and $w_\text{RX}[m,q]$ can be applied at TX and RX side, respectively. In this work we use a rectangular window function 
\be
w_\text{TX}[m,q]=w_\text{RX}[m,q]=1
\ee
for 
\be
(m,q)\in\mathcal{I}_M\times\mathcal{I}_N
\ee
and zero otherwise. 

We define the effective channel frequency response as 
\be
g[m,q]=w_\text{RX}[m,q]g'[m,q]w_\text{TX}[m,q]
\ee
resulting in the simplified expression
\be
y[m,q]=\underbrace{g[m,q]x[m,q]}_{z[m,q]}+n[m,q]\,,
\label{eq:simpleOFDMsignalmodel}
\ee
where
\be
z[m,q]=g[m,q]x[m,q]
\label{eq:TFproduct}
\ee
is the product of the transmit signal with the time- and frequency-selective channel.

\subsection{Geometry Based Channel Model}
\label{sec:GSCM}
We approximate the non-stationary fading process \cite{Bernado14} as wide-sense stationary for the duration of $M$ OFDM symbols for $m\in\mathcal{I}_M$, and $N$ subcarriers for $q\in\mathcal{I}_N$ \cite{Paier08, Renaudin10}. Hence, we model the time-variant path delay as 
\be
\tau_\ell(t)=\tau_\ell(0) - f_\ell t/\nfC
\ee
for the duration of $M \nTS$ where $f_\ell$ denotes the Doppler shift of path $\ell$, $\nfC$ the carrier frequency, and  
\be
\nTS=\nTC (N+G)
\ee
denotes the duration of an OFDM symbol.

The time-variant frequency-response $g'[m,q]$ is defined as 
\be
g'[m,q]=g_\text{TX}[q]g_\text{RX}[q]\underbrace{\sum_{\ell=0}^{P-1}\eta_\ell{\me^{-\mj2\pi \theta_\ell q}} \me^{\mj2\pi \nu_\ell m }}_{{g_\text{Ph}[m,q]}}\,,
\label{eq:GSCM}
\ee
where 
\be
\nu_\ell=f_\ell \nTS
\ee
denotes the normalized Doppler shift and 
\be
\theta_\ell=\tau_\ell(0)/(N\nTC)
\ee
the normalized path delay. The TX-filtering and RX-filtering is denoted by $g_\text{TX}[q]$ and $g_\text{RX}[q]$, respectively.

\vspace{1em}
\section{OTFS Precoding via DSFT} 
Conceptually, the DSFT converts a signal from the delay-Doppler (DD) domain to the TF-domain. The DSFT is defined as
\begin{multline}
g[m,q]=\sum_{n=0}^{M-1}\sum_{p=0}^{N-1}S_{g}[n,p]\me^{\,\mj 2\pi(mn/M-qp/N)}\\
=\text{DSFT}(S_g[n,p])\,,
\label{eq:DSFT}
\end{multline}
where $n$ and $p$ denote the indices in the Doppler and delay dimension, respectively.
The inverse discrete symplectic Fourier Transform is defined as
\begin{multline}
S_{g}[n,p]=\frac{1}{MN}\sum_{m=0}^{M-1}\sum_{q=0}^{N-1}g[m,q]\me^{-\mj 2\pi(mn/M-qp/N)}\\ =\text{IDSFT}(g[m,q])\,.
\label{eq:IDSFT}
\end{multline}

In OTFS each data symbol $d[n,p]$, $n\in\mathcal{I}_M$ and $p\in\mathcal{I}_N$ is linearly spread on the time-frequency grid 
\be
\mathcal{I}_M \times \mathcal{I}_N = \{0,\ldots,M-1\}\times\{0,\ldots,N-1\}\,,
\ee
which is defined by the OFDM physical layer. In total $MN$ orthogonal spreading functions defined by the DSFT are used. Hence, the input signal $x[m,q]$ for the OFDM modulator is calculated by transforming the information symbols $d[n,p]$ with the (symmetric) DSFT \cite{Hadani17}
\be
x[m,q] =\frac{1}{\sqrt{MN}}\sum_{n=0}^{M-1}\sum_{p=0}^{N-1}d[n,p]\me^{\mj 2\pi(mn/M-qp/N)}\,,
\label{eq:sDSFT}
\ee
where the term $\frac{1}{\sqrt{MN}}$ ensures that the energy of the information symbols does not change.

\subsection{Twisted Convolution}
The product \eqref{eq:TFproduct} defined in the TF-domain can be also expressed in the DD-domain by means of the twisted convolution \cite{Feichtinger08}
\begin{multline}
S_z[n,p]=S_{g}[n,p] \star d[n,p]= d[n,p] \star S_{g}[n,p]= \\
=\sum_{a=0}^{M-1}\sum_{b=0}^{N-1}S_{g}[a,b] \cdot \\
d[(n-a) \bmod M, (p-b) \bmod N] \me^{\mj 2\pi (n-a)b}=\\
=\sum_{a=0}^{M-1}\sum_{b=0}^{N-1}d[a,b] \cdot \\
S_{g}[(n-a) \bmod M, (p-b) \bmod N] \me^{\mj 2\pi (n-a)b}\,,
\label{eq:DDconvolution}
\end{multline}
where $\star$ denotes the twisted convolution operator. We will use this relationship later in Section \ref{sec:Detection} to develop an equalization and detection algorithm for OTFS.

\vspace{1em}
\section{MMSE Equalization}
\label{MMSE}
Assuming pilots are interleaved with data symbols in the TF grid, channel estimates $\hat{g}[m,q]$ can be obtained \cite{Zemen12, Zemen12a}. Performing minimum mean square equalization (MMSE)
\be
\hat{x}[m,q]=\frac{y[m,q]\hat{g}[m,q]^{*}}{|\hat{g}[m,q]|^2+\sigma^2}
\label{eq:MMSE} 
\ee
and de-spreading with the symmetric IDSFT provides estimates of the transmitted data symbols
\begin{multline}
\hat{d}[n,p]=\frac{1}{\sqrt{MN}}\sum_{m=0}^{M-1}\sum_{q=0}^{N-1}\hat{x}[m,q]\me^{-\mj 2\pi(mn/M-qp/N)}\\
=\sqrt{MN}\,\IDSFT(\hat{x}[m,q])
\end{multline}
After demapping $\hat{d}[n,p]$, de-interleaving and decoding, estimates of the transmitted information bits can be obtained. 
 
Implementing OTFS this way does \emph{not} provide additional diversity due to the MMSE equalization \eqref{eq:MMSE}. This is similar to the loss in spatial diversity owing to MMSE equalization of MIMO systems \cite{Mehana12}. Hence, a low-complexity maximum likelihood detection method is needed for OTFS, which we present in the next section.

\vspace{1em}
\section{Soft-Symbol Interference Cancellation and Maximum Likelihood Detection}
\label{sec:Detection}
In this paper we analyze two soft-symbol interference cancellation schemes, one in the TF-domain and another one in the DD (transform) domain.

\subsection{Time-Frequency Domain} 
For the formulation in the TF-domain we insert \eqref{eq:sDSFT} into \eqref{eq:simpleOFDMsignalmodel} and rewrite in vector-matrix notation
\be
\Vec{y}=\underbrace{\diag(\Vec{g})\Vec{S}\Vec{d}}_\Vec{z}+\Vec{n}
\ee
where vector 
\be
\Vec{y}=\left[
\begin{matrix} 
\Vec{y}[0]\\
\vdots\\
\Vec{y}[M-1]
\end{matrix}\right] \in\mathbb{C}^{MN \times 1}
\ee
stacks the vectors
\be
\Vec{y}[m]=\Big[y\left[m,0\right], \ldots, y\left[m,N-1\right]\Big]\oT
\ee
for $m\in\{0,\ldots, M-1\}$. Vectors $\Vec{g}$, $\Vec{d}$, $\Vec{n}$ and $\Vec{z}$ are defined using the same structure. We define the spreading matrix $\Vec{S}\in\mathbb{C}^{MN\times MN}$ with columns containing the vectorized basis function of the symmetric DSFT \eqref{eq:sDSFT}.

Furthermore we define the effective spreading matrix
\be
\Vec{\tilde{S}}=\diag(\Vec{g})\Vec{S}\,,
\ee
which combines the effect of linear spreading, the time- and frequency-selective channel and the  windowing at TX and RX, resulting in the expression
\be
\Vec{z}= \Vec{\tilde{S}}\Vec{d}\,.
\label{eq:TFOTFS}
\ee

Full complexity ML decoding of \eqref{eq:TFOTFS} is not possible due to the large search space of $|\mathcal{A}|^{MN}$ where $\mathcal{A}$ denotes the used symbol alphabet. Hence, we resort to an iterative ISI cancelling algorithm using soft-symbol feedback that is similar to the one presented in \cite{Zemen06c} in the context of multi-user detection. The data frame consists of $MN$ data symbols with index positions defined by the set $\Omega=\{(0,0), \ldots (M-1,N-1)\}$,  $|\Omega|=MN$.

\vspace{1em}
\subsubsection*{TF-Algorithm}
\begin{enumerate}
\item For iteration $i=1$ perform MMSE detection, see Section \ref{MMSE}.
\item For iteration $i>1$ perform ISI cancellation and ML decoding for a single data symbol $\omega\in\Omega$ using a soft-output sphere decoder \cite{Dumard08a}. Loop through all elements of the $MN$ data symbols in the set $\Omega$.

We can express parallel ISI cancellation (PIC) for the data symbol at index $\omega$ as
\be
\Vec{v}_\omega^{(i)}=\Vec{y}-\Vec{\tilde{S}}\Vec{\tilde{b}}^{(i-1)} + \Vec{\tilde{s}}_\omega\tilde{b}_\omega^{(i-1)}\,,
\ee
where superscript $\cdot^{(i)}$ denotes the iteration index and vector $\Vec{\tilde{s}}_\omega$ contains the column of $\Vec{\tilde{S}}$ for index element $\omega$.
The soft symbols $\tilde{b}_\omega$ are obtained from the extrinsic probability (EXT) output of the BCJR decoder \cite{BCJR74a} after interleaving and mapping to the used alphabet constellation $\mathcal{A}$, see \cite{Zemen06c}. Finally the ML expression for the transmitted data symbol $b_\omega$ is given by
\be
\hat{b}_\omega^{(i)}=\argmin_{b_\omega \in\mathcal{A}}\{\|\Vec{v}_\omega^{(i)} - \Vec{\tilde{s}}_\omega b_\omega \|^2\}\,.
\label{eq:TFML}
\ee 
We apply a soft-output sphere decoder \cite{Studer08} to solve \eqref{eq:TFML} and the obtained log-likelihood ratio is used as input for the BCJR decoder.
 
\item Continue with 2) until error free decoding is achieved or the maximum number of iterations $i=I$ is reached.
\end{enumerate}

\subsection{Delay-Doppler Domain}
In the DD-domain we can write (\ref{eq:DDconvolution}) in matrix-vector notation as follows 
\be
\Vec{S}_\Vec{z}= \Vec{G} \Vec{d}
\label{eq:DDOTFS}
\ee
which is the equivalent to \eqref{eq:TFOTFS} but in the DD (transform) domain. Matrix $\Vec{G}$ contains the elements of $S_g[n,p]$ for the DD-grid $\mathcal{I}_M\times\mathcal{I}_N$, arranged according to the fourth line in \eqref{eq:DDconvolution}. The DD-domain representation of $g[m,q]$ can be obtained using the IDSFT transform according to \eqref{eq:IDSFT}.

\vspace{1em}
\subsubsection*{DD-Algorithm}
\begin{enumerate}
\item Transform the received samples $y[m,q]$ to the DD-domain.
\be
Y[n,p]=\sqrt{MN}\,\IDSFT(y[m,q])
\ee

\item Perform ISI cancellation and ML decoding using a soft-output sphere decoder. Loop through all elements of the $MN$ data symbols in the set $\Omega$.

We can express parallel ISI cancellation for the data symbol at index $\omega$ as
\be
\Vec{v}_\omega^{(i)}=\Vec{Y}-\Vec{G}\Vec{\tilde{b}}^{(i-1)} + \Vec{g}_\omega\tilde{b}^{(i-1)}_\omega\,.
\ee
The ML expression for the transmitted data symbol $b_\omega$ is given by
\be
\hat{b}_\omega^{(i)}=\argmin_{b_\omega\in\mathcal{A}}\{\|\Vec{v}_\omega^{(i)} - \Vec{g}_\omega b_\omega\|^2\}\,,
\label{eq:DDML}
\ee 
where vector $\Vec{g}_\omega$ contains the column of $\Vec{G}$ for index element $\omega$.
We use a soft-symbol output sphere decoder \cite{Studer08} for the efficient solution of \eqref{eq:DDML}.
\item Continue with 2) until error free decoding or the maximum number of iterations $i=I$ is reached.
\end{enumerate}

\vspace{1em}
\section{Numerical Simulation Results}
For the numerical link level simulations we use a channel model implemented as shown in Sec. \ref{sec:GSCM}. The channel model uses an exponentially decaying power-delay profile (PDP) with root-mean square delay spread of $0.4\,\mu\text{s}$ and a Clarke Doppler power spectral density \cite{Clarke68} according to the velocity $v\in\{0,200\}\,\text{km/h}$ for a carrier frequency of $\nfC=5.9\,\text{GHz}$. 

A convolutional channel code with rate $R=1/2$ and code polynomial $(5,7)_8$ is used for channel coding after random interleaving of the data bits. The OFDM system uses $N=64$ subcarriers, a cyclic prefix with length $G=16$, and a frame length of $M=36$ OFDM symbols. The bandwidth $B=10\,\text{MHz}$. These parameters resemble the physical layer (PHY) of IEEE 802.11p. For all simulation we assume that prefect channel state information (CSI) is available at the RX-side\footnote{CSI estimates can be obtained from pilot information multiplexed with data symbols in the DD-domain or the TF-domain \cite{Rakib16}.}.

\begin{figure}
	\centering
	\includegraphics[width=\columnwidth]{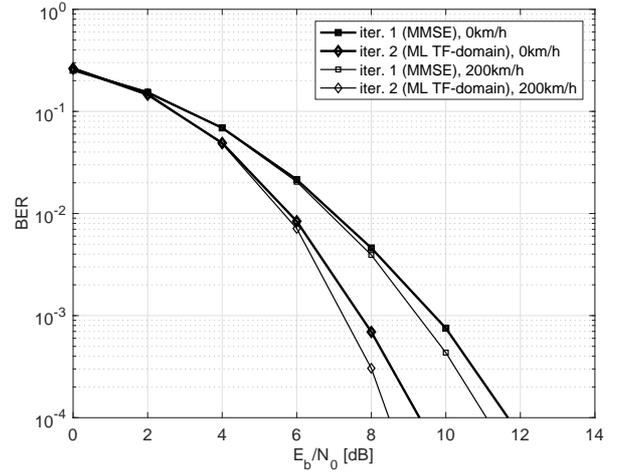}
	\caption{BER versus $\nEb/N_0$ for equalization in the TF-domain for velocities $v\in\{0,200\}\,\text{km/h}$.}
	\label{fig:TFdec}
\end{figure}

In Fig. \ref{fig:TFdec} we show the bit error rate (BER) versus $\nEb/N_0$.  
\begin{figure}
	\centering
	\includegraphics[width=\columnwidth]{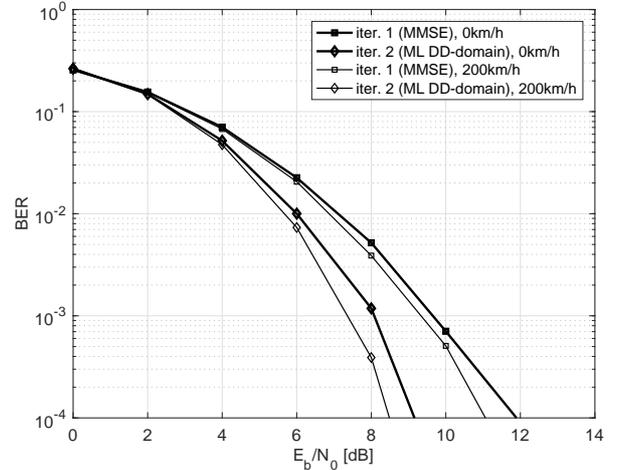}
	\caption{BER versus $\nEb/N_0$ for equalization in the DD-domain.}
	\label{fig:DDdec}
\end{figure} 
The first iteration employs MMSE equalization, which can be implemented with small numerical complexity but lacks full diversity. The results of the second iteration are obtained with soft-symbol ISI cancellation \cite{Zemen06c} and (near) ML decoding with a soft-output sphere decoder \cite{Studer08} implemented in the TF-domain. Figure \ref{fig:TFdec} shows a gain of about $2.5\,\text{dB}$ for a BER of $10^{-4}$ for a velocity of $v=0\,\text{km/h}$ from iteration one to iteration two. The second set of results for $v=200\,\text{km/h}$ show a steeper slope due to additional time (Doppler) diversity. The gain for the second iteration is similar.

In Fig. \ref{fig:DDdec} we show results for equalPization in the DD (transform) domain. The obtained performance results are extremely similar to the one shown in Fig. \ref{fig:TFdec} for the TF-domain. This can be explained by the fact that the TF and the DD-domain are related by a linear transform, hence, leading to similar results.

\begin{figure}
	\centering
	\includegraphics[width=\columnwidth]	{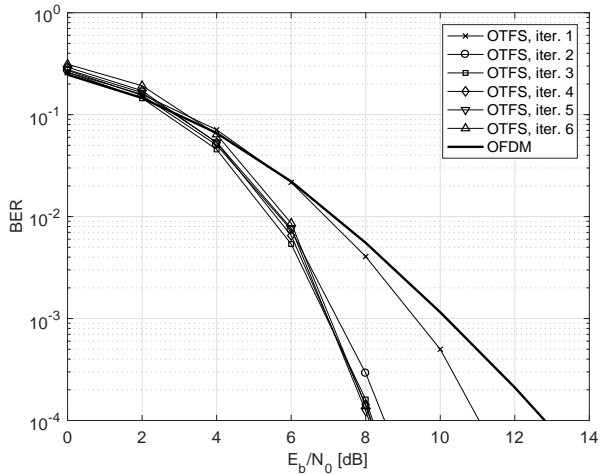}
	\caption{BER versus $\nEb/N_0$ comparing OFDM with OTFS for different iterations for $v=200\,\text{km/h}$.}
	\label{fig:OFDMvsOTFS}
\end{figure}

In Fig. \ref{fig:OFDMvsOTFS} we analyze the performance of OTFS vs. the number of iterations and we provide a comparison with OFDM, i.e. using no precoding with OTFS. The parameters are kept similar as before and the results are shown for a velocity of $v=200\,\text{km/h}$.

In the first iteration OTFS uses MMSE equalization and we can see that only a small diversity gain compared to OFDM can be obtained at higher SNR. The largest gain is obtained in the second iteration where PIC is used to suppress ISI. We implement ML decoding with a soft-output sphere decoder. The third iteration provides an additional small gain. No further improvement is obtained for iterations four to six.

\vspace{1em}
\section{Conclusions}
In this paper we described an equalization and detection scheme for OTFS that allows to obtain the full diversity of a time- and frequency-selective communication channel. This scheme uses MMSE equalization in the first iteration and PIC with a soft-output sphere decoder from the second iteration onwards. A maximum of three iteration was shown to provide BER improvements. We can demonstrate a gain of $5\,\text{dB}$ comparing OTFS with OFDM for a BER of $10^{-4}$ and a relative velocity of $200\,\text{km/h}$.

\vspace{1em}
\section*{Acknowledgment}
We would like to thank Olivier Renaudin for helpful discussions and comments.

\vspace{1em}

%\bibliography{/Users/zement/Documents/Publications/IEEEabrv,/Users/zement/Documents/Publications/MasterBibZemen}

\end{document}